\pdfoutput=1
\documentclass[sigconf]{acmart}
\renewcommand\footnotetextcopyrightpermission[1]{} 
\usepackage{xcolor}
\usepackage{tikz}
\usepackage{pgfplots}
\pgfplotsset{compat=newest}
\usepackage{booktabs}
\AtBeginDocument{%
  \providecommand\BibTeX{{%
    \normalfont B\kern-0.5em{\scshape i\kern-0.25em b}\kern-0.8em\TeX}}}

\copyrightyear{2020}
\acmYear{2020}
\setcopyright{rightsretained}

\begin{document}

\title{TwinBERT: Distilling Knowledge to Twin-Structured BERT Models for Efficient Retrieval}

\author{Wenhao Lu}
\affiliation{%
  \institution{Bing Ads of AI \& Research Group}
  \institution{Microsoft}
  \streetaddress{One Microsoft Way}
  \city{Redmond}
  \state{WA} 
  \postcode{98052-6399}
}
\email{wenhlu@microsoft.com}

\author{Jian Jiao}
\affiliation{%
  \institution{Bing Ads of AI \& Research Group}
  \institution{Microsoft}
  \streetaddress{One Microsoft Way}
  \city{Redmond}
  \state{WA} 
  \postcode{98052-6399}
}
\email{jiajia@microsoft.com}

\author{Ruofei Zhang}
\affiliation{%
  \institution{Bing Ads of AI \& Research Group}
  \institution{Microsoft}
  \streetaddress{One Microsoft Way}
  \city{Redmond}
  \state{WA} 
  \postcode{98052-6399}
}
\email{bzhang@microsoft.com}



\begin{abstract}
Pre-trained language models like BERT have achieved great success in a wide variety of NLP tasks, while the superior performance comes with high demand in computational resources, which hinders the application in low-latency IR systems. We present TwinBERT model for effective and efficient retrieval, which has twin-structured BERT-like encoders to represent query and document respectively and a crossing layer to combine the embeddings and produce a similarity score. Different from BERT, where the two input sentences are concatenated and encoded together, TwinBERT decouples them during encoding and produces the embeddings for query and document independently, which allows document embeddings to be pre-computed offline and cached in memory. Thereupon, the computation left for run-time is from the query encoding and query-document crossing only. This single change can save large amount of computation time and resources, and therefore significantly improve serving efficiency. Moreover, a few well-designed network layers and training strategies are proposed to further reduce computational cost while at the same time keep the performance as remarkable as BERT model. Lastly, we develop two versions of TwinBERT for retrieval and relevance tasks correspondingly, and both of them achieve close or on-par performance to BERT-Base model.

The model was trained following the teacher-student framework and evaluated with data from one of the major search engines. Experimental results showed that the inference time was significantly reduced and was firstly controlled around 20ms on CPUs while at the same time the performance gain from fine-tuned BERT-Base model was mostly retained. Integration of the models into production systems also demonstrated remarkable improvements on relevance metrics with negligible influence on latency.
\end{abstract}


\keywords{Deep Learning; Deep Neural Network (DNN); Semantic Embedding; Information Retrieval; k-NN; CDSSM; PyTorch; Sponsored Search; BERT; Knowledge Distillation}


\maketitle

\fancyhf{} 
\renewcommand{\headrulewidth}{0pt}

\section{Introduction}

Pre-trained language models such as BERT \cite{devlin2018bert} and GPT  \cite{radford2018improving} have led a series of breakthroughs in a broad variety of NLP tasks including question answering, natural language inference, sentiment classification and others, and more impressively, they even surpassed human performance on some of them \footnote{\url{https://gluebenchmark.com/leaderboard/}}. However, to serve these deep-structured models in a production system, besides accuracy, latency is also an important factor to consider. A BERT-Base model, for instance, has 110 million parameters and 12 stacked multi-head attention networks, which is extremely computationally intensive and makes it challenging to deploy such a model in a real-world system.

In the age of information explosion, to meet people's information needs, a variety of modern applications have been developed including web search, online advertising, product recommendation, digital assistant and personalized feed. At the heart of these systems, information retrieval (IR) plays an important role in handling the increasingly growing volume of information. The quality of an IR system crucially depends on the deep understanding of queries and documents, which fundamentally is an NLP problem and could benefit from the state-of-the-art pre-trained models. However, considering the large-scale and low-latency nature of IR systems, the long inference time of these models becomes a bottleneck for their applications in the area. Most of the prior knowledge distillation efforts on BERT \cite{jiao2019tinybert}, \cite{mirzadeh2019improved}, \cite{sun2019patient}, \cite{tang2019distilling} showed effectiveness in compressing the complex models and reducing the inference time to a certain degree. Nevertheless, few of them could meet the latency requirement of IR systems. 

To address the latency problem brought by the advanced NLP techniques, this paper proposes a novel language representation model for efficient retrieval, called TwinBERT. The model has twin-structured BERT-like encoders to encode the query and document respectively and a crossing layer to combine the two embeddings and produce a similarity score. The model was evaluated using data collected from one of the major search engines and the accuracy performance is close to a complex BERT-Base model. More importantly, the implementation with PyTorch, although not as efficient as C/C++ and others, showed considerable reduction in inference time, and the average time cost on CPU over 1,000 random queries was only around 20 milliseconds for scoring 100 documents.  

Our contributions are summarized as below.
1) a twin-BERT structure which separates BERT inputs when encoding and allows embeddings to be pre-computed 2) an efficient retrieval model based on cosine similarity supporting ANN search 3) an efficient relevance prediction model based on residual network with performance close to BERT-Base. 

The rest of the paper is organized as follows. Section \ref{sec:rev} presents a literature review on related works. Section \ref{sec:adv} briefly introduces the context of paid search. Section \ref{sec:TB} discusses details of TwinBERT including network architecture, model training and online serving. Section \ref{sec:exp} reports the experimental results of TwinBERT compared to baseline models. Section \ref{sec:prod} introduces how TwinBERT is deployed and used in production system. In the end, Section \ref{sec:con} concludes the work and lists future directions to explore.

\section{Related Work}
\label{sec:rev}
\textbf{Learning Representations through Language Models}\space\space\space

Language representations, as the building blocks of NLP models, are impressively effective in improving model performance on NLP tasks, and therefore have become an important research area over the years. According to how the representations are employed in downstream tasks, prior works in the area can be broadly grouped into two categories: feature-based approaches and fine-tuning approaches. Word representations, sentence-based representations, and most recently contextual word representations are three directions of the feature-based representations. Word2Vec \cite{mikolov2013efficient}, GloVe \cite{pennington2014glove} and FastText \cite{bojanowski2017enriching} focus on learning word representations and different senses of a word are all combined into one vector while Skip-thought \cite{kiros2015skip}, FastSent \cite{hill2016fastsent}, Quick-thought \cite{laja2018quick}, Universal sentence encoder \cite{daniel2018use} and other works \cite{conneau2017supervised}, \cite{subra2018multitask} extract sentence-level representations. Unlike the previous works, ELMo \cite{peters2018deep} derives word representation based on the entire sentence and captures representations of words on multiple granularity by combining vectors from intermediate layers of a multi-layer BiLSTM. All these methods only require one round of training before used in any downstream tasks. In recent two years, pre-trained models such as GPT \cite{radford2018improving} and BERT \cite{devlin2018bert} demonstrated superior performance. In contrast to previous works, the representations from these models are learned in two phases. In the first phase, a language model is learned in an unsupervised manner. In the second phase, the model is fine-tuned with task-specific label data to produce representations used in downstream tasks.

BERT stands for bidirectional encoder representations from transformers and achieved the state-of-the-art performance on a broad variety of NLP tasks. BERT is pre-trained on a large corpus of unlabelled text data including the entire Wikipedia and BooksCorpus \cite{bookscorpus}. It has two pre-training tasks: masked language model (MLM) and next sentence prediction (NSP). MLM enforces the model to learn parameters by optimizing the prediction of masked tokens. To better serve the downstream binary classification tasks such as question answering (QA) and natural language inference (NLI), NSP is introduced to jointly train with MLM, which requires a pair of sentences as input. Through the multi-layer bidirectional structure, tokens from the two sentences deeply interact with each other and as a result, model performance is effectively improved for binary classification tasks. As a side effect, the computational cost is also highly increased, especially in the area of information retrieval, where one query needs to be paired with a large number of document candidates. BERT has overwhelming influence and to extend the work, a few variants have been developed including MTDNN \cite{liu2019multi}, XLNet \cite{yang1906xlnet}, ERNIE \cite{zhang2019ernie}, RoBERTa \cite{liu2019roberta}, ALBERT \cite{lan2019albert} and T5 \cite{raffel2019exploring}.
\\
\\
\noindent\textbf{Distilling Knowledge to Compact Models}\space\space\space

With limited computational resources and strict latency requirement, expensive models such as BERT normally cannot be directly deployed in real-world applications, and knowledge distillation (KD) \cite{bucilua2006model},\cite{hinton2015distilling} is typically adopted to address the issue. The idea is to transfer the knowledge learnt from an expensive high-performance teacher model to a compact student model without significant performance loss. In contrast to traditional machine learning tasks, a loss function is defined on soften probabilities produced by the teacher model instead of hard labels, which are so-called soft labels, and soft labels supposedly have higher entropy which could provide more information and less variance. Prior efforts in KD on deep-structured models like BERT mainly focused on the transfer techniques. \cite{tang2019distilling} augmented the training data for distillation with synthetic examples.  BERT-PKD \cite{sun2019patient} learned distilled knowledge from intermediate layers besides the output layer. \cite{turc2019well} demonstrated pre-trained student had better performance than random initialization. TinyBERT \cite{jiao2019tinybert} further expanded distillation to transformer layers. \cite{mirzadeh2019improved} proposed teacher assistant to bridge the gap between student and teacher. MT-DNN ensemble \cite{liu2019improving}, \cite{zhu2019panlp} and MKDN \cite{yang2019model} improved the student model performance via learning from multiple teachers.

To the best of our knowledge, none of these works have attempted to decouple the two-sentence input, which could therefore reduce the inference time complexity for two-input cases from quadratic to linear time complexity. 

\section{Sponsored Search}
\label{sec:adv}
TwinBERT is developed in the context of sponsored search. Readers can refer to \cite{edelman2007internet} for a comprehensive introduction of the topic. In short, sponsored search engine delivers ads alongside the organic search results. There are often three parties involved in the sponsored search ecosystem: the user, the advertiser and the search platform. The goal of the platform is to display a list of ads that best match user's intent. Below is the minimum set of
key concepts for discussions that follow.

\begin{description}
    \item[Query:] A short text string that contains user's intent. Users enter queries in a search box to look for related information.
	\item[Keyword:] A short text string that expresses advertiser's intent. Keywords are provided by advertisers and are not visible to end users but they are pivotal in that search engine relies on them to match user intents.
	\item[Impression:] An ad being displayed to the end user, on the result page of the search engine.
\end{description}

On the backend of a paid search engine, the number of keywords created by advertisers are typically at the scale of billions. Fast IR techniques are firstly applied to reduce the number of keywords to a much smaller matched subset and then sent to downstream components, where more complex and less efficient algorithms are used to finalize the ads to display.

To be consistent with the above context, keywords are used instead of documents throughout the paper. 

\section{TwinBERT}
\label{sec:TB}

The architecture of TwinBERT is presented in this section with a few well-designed network layers to balance the effectiveness and efficiency. Other topics including model training and online serving are also discussed in detail.

\subsection{Model Architecture}

\begin{figure}
	\centering
		\includegraphics[scale=0.5]{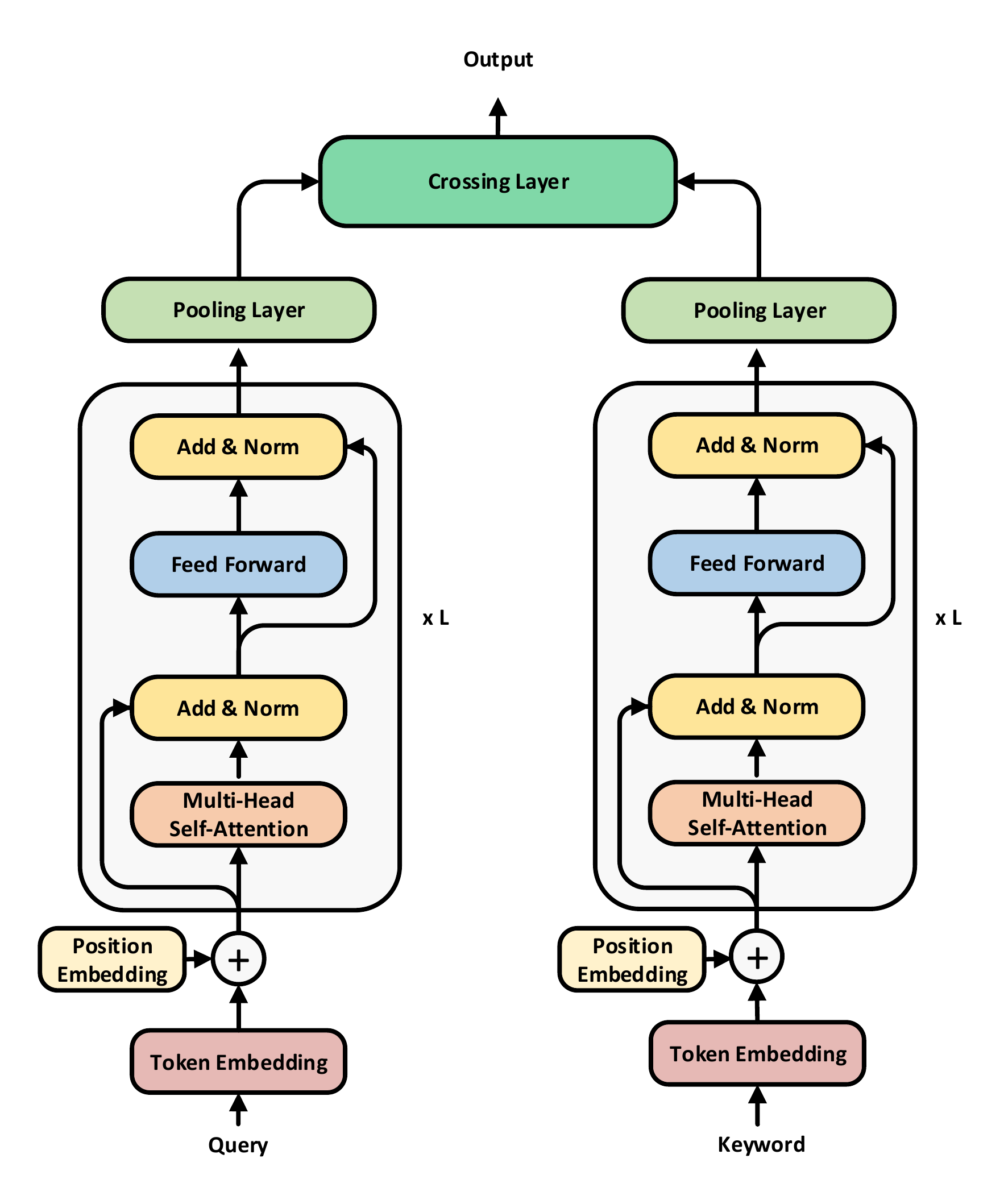}
	\caption{TwinBERT Architecture}
	\label{fig:arch}
\end{figure}
As shown in Figure \ref{fig:arch}, the architecture of TwinBERT consists of two multi-layer transformer encoders and a crossing layer to combine the vector outputs of encoders and produce the final output. It is noteworthy that the parameters of the two encoders of query and keyword could be shared or different. The detailed comparison of the two styles is discussed in Section \ref{sec:exp}.

Similar to BERT model architecture, at the bottom of each encoder is the embedding layer, where the query and keyword sentences are represented separately as embeddings and then fed into corresponding encoders. The middle part of each encoder is a stack of transformer encoders with the same implementation as described in \cite{vaswani2017attention} but a different setting. Following the notations in BERT, the number of layers is denoted as $L$, the hidden size is $H$, and the number of self-attention heads is $A$. In this work, the performance is mainly reported with the following model setting: $L=6$, $H=512$ and $A=8$ (the size of the feed-forward intermediate layer is also set to equal to $H$). The last and top layer of the encoder is the weighted pooling layer which applies a weighted sum of the final hidden vectors and produces a single embedding for each input sentence.

\subsection{Input Representation}
In TwinBERT, the two input sentences are decoupled and encoded separately, with each encoder only taking care of one single sentence. Different from BERT, there is no need to introduce a separator token $[SEP]$ to separate the two segments and the input sequence length is roughly reduced by half. According to \cite{vaswani2017attention}, the per-layer complexity of self-attention is $O(n^2)$ on sequence length and other operations are $O(n)$. As a result of the cut on sequence length, the overall inference cost is correspondingly decreased. The other classification token $[CLS]$ in BERT is dropped in weighted-average pooling while reserved only in classification token pooling, which will be discussed in the pooling layer section. 

For token embeddings, TwinBERT uses the tri-letter based word embeddings introduced in \cite{shen2014learning}. Compared to the 30K dimensional WordPiece embeddings \cite{wu2016google} in BERT, tri-letter based embeddings have larger vocabulary size (50K), and therefore can bear more information for better performance. On the other hand, they are more efficient when extracted at inference time since the extraction of each token is independent, while WordPiece extraction is a recursive process.

BERT embeddings are combinations of three components: token embeddings, segment embeddings and position embeddings. While, the input of a TwinBERT encoder only contains one single sentence and segment embeddings are unnecessary. Therefore, the input embeddings only consist of the sum of token embeddings and position embeddings.
\subsection{Pooling Layer}
The output of the encoder is a sequence of vectors, each corresponding to an input token with position information implied from its index in the input sentence. To provide a unique fix-length vector representation for both inputs, a pooling layer is added to provide a robust approach to unify all token vectors into a single sentence level embedding. Specifically, two pooling methods are experimented: weighted-average pooling and classification token pooling. Compared to standard average pooing, {\bfseries weighted-average pooling} introduces a weight to each token vector and the output is the weighted-average of all token vectors. The weight parameters are learned as part of the entire network. The second method is inspired by the special classification token ($[CLS]$) in BERT and is so called {\bfseries classification token pooling}. The implementation involves prefixing the sequence with $[CLS]$ at the input layer. The output of encoder is simply the final hidden vector of $[CLS]$. Comparison results of the two methods are presented in the experiment section.

\subsection{Crossing Layer}
Given the sentence embeddings of query and keyword, here comes the question: how to combine the two? Two versions of TwinBERT are proposed to address the problem, denoted as TwinBERT$_{\textrm{cos}}$ and TwinBERT$_{\textrm{res}}$ respectively:

\noindent\textbf{Cosine similarity}\space\space\space Cosine similarity is an intuitive approach for combining two vectors of the same length. Formally, cosine similarity is defined as
\begin{equation}
    \cos(\mathbf{q}, \mathbf{k})=\frac{\mathbf{q}\cdot\mathbf{k}}{\|\mathbf{q}\|\cdot\|\mathbf{k}\|}
\end{equation}
, where $\mathbf{q}$ and $\mathbf{k}$ correspond to the embedding vectors of query and keyword. The output falls in the range of $[-1,1]$, while the soft targets from teacher model are between 0 and 1. In order to align the two, an additional logistic regression layer is applied to the cosine similarity score and convert it to $[0,1]$. 

Cosine similarity projects the two embeddings to the same vector space, and when both vectors are normalized, it can be easily transformed into Euclidean distance. Thereupon, approximate nearest neighbor (ANN) algorithms can be naturally applied for retrieval tasks \cite{huang2013learning}.
	
\noindent\textbf{Residual network}\space\space\space Residual networks were firstly proposed in \cite{he2016deep} to solve the image recognition problem. Inspired by \cite{shan2016deep}, where residual layers were used in a non-convolutional network in the NLP domain, they are adopted here to overcome over-fitting and gradient vanishing problems. Specifically, the embeddings of query and keyword are first combined by a max operator and then fed into the residual connection. The formal definition for residual function is as follows:
\begin{equation}
    \mathbf{y}=\mathcal{F}(\mathbf{x}, W, b) + \mathbf{x}
\end{equation}
, where $\mathbf{x}$ is the max of query vector $\mathbf{q}$ and keyword vector $\mathbf{k}$ and $\mathcal{F}$ is the mapping function from $\mathbf{x}$ to the residual with parameters $W$ and $b$. Using concatenation instead of max operator is another option. Here, the motivation behind choosing max over concatenation is that it provides a down-sampling effect and also softly maps the two embeddings to a closer vector space. 

Similarly, a logistic regression layer is applied to the output vector of residual function $\mathbf{y}$ to predict the binary relevance label. Compared to cosine similarity, the deep-structured network could model more complex problems and therefore produce better performance, but as a tradeoff, it is less efficient in computation and cannot easily work with the ANN algorithms.

\subsection{Knowledge Distillation}

TwinBERT is trained following the teacher-student framework via knowledge distillation since comparing to learning from scratch, student models usually have better performance \cite{hinton2015distilling}. For simplicity, Google's 12-layer BERT-Base model is fine-tuned using editorial query-keyword relevance labels as the teacher model and is then used to score a collection of impressed query-keyword pairs. The logits $z$ are outputted to generate soft labels using the following equation
\begin{equation*}
  y_i = \frac{\exp(z_i/T)}{\sum_j \exp(z_j/T)}
\end{equation*}
, where $T$ is the temperature parameter controlling softness of the labels. When $T=1$, it is equivalent to standard softmax function. As $T$ grows, the target values become softer and hence provide more information. Specifically, in TwinBERT, $T$ is set to 2.

The cross-entropy loss function for binary classification is defined as
\begin{equation*}
loss=-\sum_{i=1}^N(y_i\log(p_i)+(1-y_i)\log(1-p_i))
\end{equation*}
, where $N$ is the number of samples and $p$ is the predicted probability. It was claimed in \cite{tang2019distilling} that mean squared error (MSE) produced better results but our experiments showed the opposite.

\subsection{Online Serving}
\label{sec:online}

\begin{figure}
	\centering
		\includegraphics[scale=0.8]{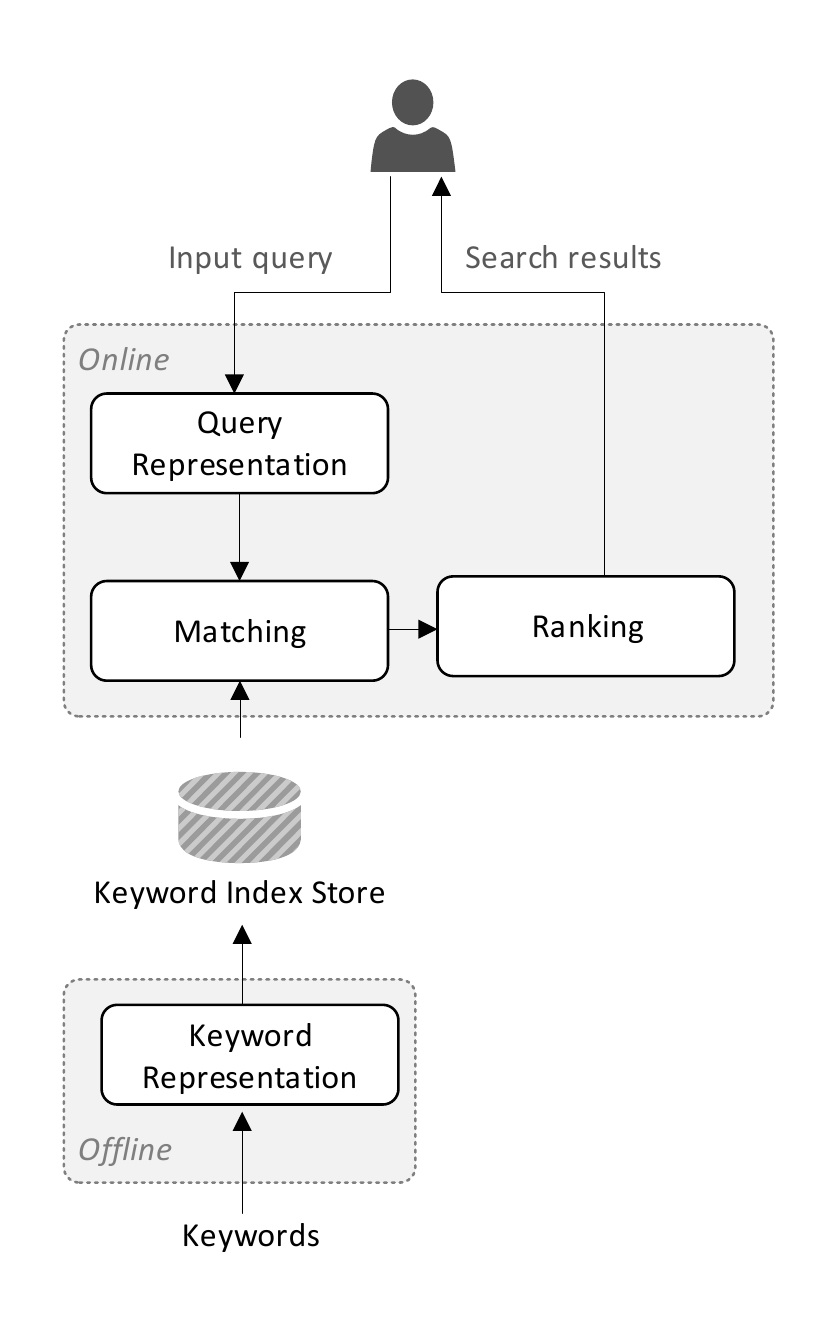}
	\caption{Online Serving}
	\label{fig:online}
\end{figure}
TwinBERT is designed to adopt the latest NLP breakthroughs to IR systems, particularly the paid search engine. Figure \ref{fig:online} outlines the high-level architecture of TwinBERT-based information retrieval system. 

The keywords that advertisers entered are stored in a distributed database. As the offline process of the system, keywords are extracted from the database and represented as embeddings $\{\mathbf{k}_j|j < m\}$ through keyword-side TwinBERT encoder. For efficient retrieval, ANN techniques such as locality-sensitive hashing \cite{datar2004locality} and k-d trees \cite{friedman1976algorithm} are typically employed to improve search performance. Specifically in TwinBERT, the keyword embeddings are stored and organized in a graph structure as described in \cite{wang2012query}.

At run-time, when a user enters a search query, the query embedding $\textbf{q}$ is generated on-the-fly by the query-side TwinBERT encoder and ANN search is performed to find the top results from the pre-built keyword indices, which are normally pre-loaded in memory.

\section{Experiments}
\label{sec:exp}
This section presents training details and experimental results of TwinBERT models. Section \ref{sec:trainingdetails} introduces the data and hyper-parameters used in teacher and student model training. Sections \ref{sec:relevanceeval} and \ref{sec:retrievaleval} give evaluation results on relevance and retrieval tasks. In the relevance experiment, a few baseline models and two versions of TwinBERT models, TwinBERT$_{\textrm{cos}}$ and TwinBERT$_{\textrm{res}}$ which differ in the design of the crossing layer, were evaluated. In the retrieval experiment, where ANN was used when searching keywords in a vector space, TwinBERT$_{\textrm{cos}}$ was picked and compared with C-DSSM \cite{shen2014learning}. In \ref{sec:trainingstrategies}, a few effective training strategies are discussed. \ref{sec:overallresults} gives the overall evaluation results. Lastly, in \ref{sec:inferencetime}, inference time of TwinBERT models with different configurations are reported for better understanding on the design of TwinBERT models.

\subsection{Teacher and Student Model Training}
\label{sec:trainingdetails}
\subsubsection{Training teacher model}
The teacher model used in this paper was a BERT-Base (BERT$_{12}$) model fine-tuned from the uncased checkpoint trained and released by the authors of \cite{devlin2018bert}. 5.8 million query-keyword data was used for fine-tuning. In the data, query-keywords were given labels which indicate 4 different levels of relevance: bad, fair, good and excellent. In the fine-tuning process, fair, good and excellent were mapped into one level which was non-bad and the model learnt query-keyword relevance from binary labels (bad vs. non-bad) based on cross-entropy loss. Hyper-parameters of fine-tuning were the same as what suggested in \cite{devlin2018bert}. Batch size was set to $2,048$ and model was trained for $5$ epochs.

\subsubsection{Training student models}
The student models in this paper were distilled from the same teacher model. 500 million impressions were sampled from log and scored by the teacher model to generate soft targets for student model training. In the training process of TwinBERT, model parameters were randomly initialized and Adam \cite{kingma2014method} was used for optimization. Training was done on four V100 GPUs and hyper-parameters were adopted from BERT pre-training \cite{devlin2018bert}: learning rate = $1e-4$, $\beta_1$ = $0.9$, $\beta_2$ = $0.999$, L2 weight decay = $0.01$. The model was trained for $10$ epochs with batch size set to $2,048$. The two encoders in TwinBERT models were trained with shared parameters.

\subsection{Evaluation on Relevance Task}
\label{sec:relevanceeval}
\subsubsection{Experiment setup}
In the relevance experiment, C-DSSM and 3-layer BERT (BERT$_3$) were chosen for baseline student models. The former proved to be effective for information retrieval tasks and here for fair comparison, the hidden size was set to be the same as what used in TwinBERT, which was 512. The latter, as a student model, was used in multiple knowledge distillation works \cite{sun2019patient}, \cite{yang2019model}. In addition, BERT$_3$ has about 46 million parameters, which is comparable to TwinBERT model (35 million). 

To evaluate the relevance performance of teacher and student models, two test sets were sampled from logs in a major sponsored search system. There are roughly 600,000 and 700,000 instances in test set 1 and 2 respectively. The two sets were sampled from different components in the system, so describe different perspectives of query-keyword relevance in the system. Both test sets are held-out.

\subsubsection{Effects of Design Choices}
In the design of TwinBERT, a few choices were experimented before the model was finalized. The results are summarized in Table \ref{tab:setting}.

\textbf{Number of layers:} Comparing Setting 2 vs. Setting 3 suggests that reducing the number of layers by half results in around 0.16\% drop in performance, which is significant when talking about hundreds of millions of impressions. If latency capacity allows, it is better to have deeper structure.

\textbf{Crossing layer:} Using max to combine the query and keyword embeddings has better performance than the naive concatenation with about 0.15\% gain (Setting 2 vs. Setting 4). Again, it is significant considering the scale of search. Max function produces an abstraction of the two representations of query and keyword, and helps with the generalization.

\textbf{Token embedding:} Character-level trigram representation outperforms WordPiece by 0.26\% (Setting 2 vs. Setting 5). Compared to WordPiece, trigrams could map different forms of a word to a similar representation and have more dimensions. In the context of sponsored search where query and keyword tend to have more out of vocabulary words (e.g., typo words or invented names), these pros are shown to be effective in boosting the performance. Besides, character-level trigrams are more efficient at extraction.

\textbf{Position embedding:} In sponsored search, both the query and keyword are often short phrases but the order of words is still important and meaningful for understanding. Position embedding helps to improve the performance by about 0.23\% by comparing Setting 2 to Setting 6.

\textbf{Classification token:} Although in BERT, the signal from classification token hidden vector proves to be effective in many downstream tasks, it is less effective than weighted average when there is a need to combine two embeddings. The difference between Setting 7 and Setting 2 is as high as 1\%, more significant than other changes.

\begin{table}
  \caption{ROC-AUC of different settings on test set 1}
  \label{tab:setting}
  \begin{tabular}{lllllll}
    \toprule
    & Token. & Pos. & Pooling & Crossing & $L$ & AUC$_1$ \\
    \midrule
    1 & Tri-letter & $\surd$ & Weighted & Cos & 6 & 0.8883 \\
    2 & Tri-letter & $\surd$ & Weighted & Max + Res & 6 & 0.9010 \\
    3 & Tri-letter & $\surd$ & Weighted & Max + Res & 3 & 0.8994 \\
    4 & Tri-letter & $\surd$ & Weighted & Concat + Res & 6 & 0.8995 \\
    5 & WordPiece & $\surd$ & Weighted & Max + Res & 6 & 0.8987 \\
    6 & Tri-letter &   & Weighted & Max + Res & 6 & 0.8989 \\
    7 & Tri-letter & $ \surd$   & CLS & Max + Res & 6 & 0.8897 \\
  \bottomrule
\end{tabular}
\end{table}

\subsubsection{Model accuracy}

Table \ref{tab:overall} shows the ROC-AUC of TwinBERT models comparing with C-DSSM, BERT$_{3}$ and BERT$_{12}$. The AUC comparison of different models is consistent on two test sets. First of all, both TwinBERT$_{\textrm{cos}}$ and TwinBERT$_{\textrm{res}}$ outperform C-DSSM model by 1.9\% and 3.4\% on test set 1 while 2.0\% and 6.3\% on test set 2, which exhibits the advance of TwinBERT model's architecture. However, the performance gap between TwinBERT$_{\textrm{cos}}$ and TwinBERT$_{\textrm{res}}$ suggests that the current cosine version is still not effective enough to express the interaction between query and keyword but the more complex residual network can. Compared with BERT$_{3}$, TwinBERT$_{\textrm{res}}$ achieves higher AUC (+0.17\% and +0.07\%), and most impressively, its performance is close to BERT$_{12}$ with only -0.01\% and -0.26\% differences, which proves the effectiveness of TwinBERT model in distilling knowledge from a BERT-like teacher model.

\begin{table}
  \caption{ROC-AUC of TwinBERT models comparing with C-DSSM, BERT$_{3}$ and BERT$_{12}$ on two test sets}
  \label{tab:overall}
  \begin{tabular}{lll}
    \toprule
    Model & AUC$_1$ & AUC$_2$\\
    \midrule
    C-DSSM & 0.8713 & 0.8571 \\
    BERT$_3$ & 0.8995 & 0.9107 \\
    TwinBERT$_{\textrm{cos}}$ & 0.8883 & 0.8743 \\
    TwinBERT$_{\textrm{res}}$ & 0.9010 & 0.9113 \\
    BERT$_{12}$ & 0.9011 & 0.9137\\
  \bottomrule
\end{tabular}
\end{table}

\subsection{Evaluation on Retrieval Task}
\label{sec:retrievaleval}
\subsubsection{Experiment setup}
In the retrieval experiment, C-DSSM was selected as the baseline and compared with TwinBERT$_{\textrm{cos}}$. Both models were trained on the same training data with the same hyper-parameters as described in the relevance experiment. The evaluation was conducted in three steps. Firstly, embeddings of queries and keywords were generated with the model, and a keyword index was built based on the embeddings of keywords. Secondly, ANN search was performed to find the top results from the pre-built keyword indices. Lastly, top N results for each query were collected and nDCG (normalized Discounted Cumulative Gain) was evaluated for each model based on the editorial labels. This time, all 4 labels were used for evaluation. In the experiment, the query set had 2,000 randomly sampled queries and keyword set had 100 million randomly sampled keywords. Top 5 results were collected for nDCG evaluation.

\begin{table}
  \caption{Density differences of all 4 labels by comparing top 5 results from TwinBERT$_{\textrm{cos}}$ and C-DSSM}
  \label{tab:retrieval}
  \begin{tabular}{llll}
    \toprule
    bad & fair & good & excellent\\
    \midrule
    -7.4\% & -2.6\% & 1.9\% & 18.8\%\\
  \bottomrule
\end{tabular}
\end{table}

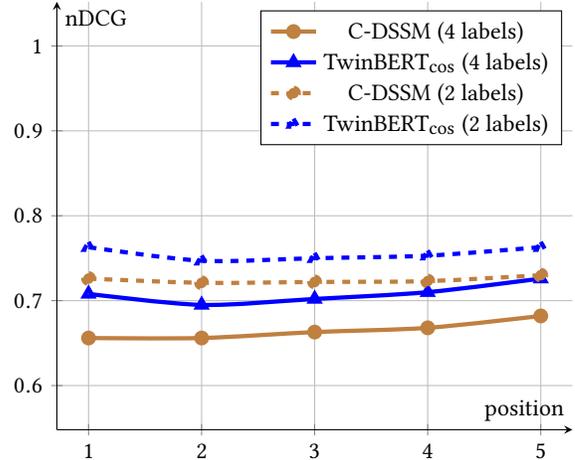
\begin{figure}
\begin{tikzpicture}
  \begin{axis}[
    grid = major,
    clip = true,
    clip mode=individual,
    restrict y to domain=0.6:1,
    axis x line = middle,
    axis y line = middle,
    xlabel={position},
    ylabel={nDCG},
    domain = 0.01:5,
    xmin = 1,
    xmax = 5,
    enlarge y limits={rel=0.13},
    enlarge x limits={rel=0.07},
    ymin = 0.6,
    ymax = 1,
    legend entries={C-DSSM (4 labels),TwinBERT$_{\textrm{cos}}$ (4 labels),C-DSSM (2 labels),TwinBERT$_{\textrm{cos}}$ (2 labels)},
  ]

    \addplot[color=brown,mark=*,samples=100,smooth,ultra thick] coordinates {
		(1,0.656)
		(2,0.656)
		(3,0.663)
		(4,0.668)
		(5,0.682)};
    \addplot[color=blue,mark=triangle,samples=100,smooth,ultra thick] coordinates {
		(1,0.708)
		(2,0.695)
		(3,0.702)
		(4,0.710)
		(5,0.726)};
		
	\addplot[color=brown,mark=*,samples=100,smooth,dashed,ultra thick] coordinates {
		(1,0.726)
		(2,0.721)
		(3,0.722)
		(4,0.723)
		(5,0.730)};

    \addplot[color=blue,mark=triangle,samples=100,smooth,dashed,ultra thick] coordinates {
		(1,0.763)
		(2,0.747)
		(3,0.750)
		(4,0.753)
		(5,0.763)};
  \end{axis}
\end{tikzpicture}
\caption{nDCGs of TwinBERT$_{\textrm{cos}}$ and C-DSSM}
\label{fig:retrieval}
\end{figure}

\subsubsection{Model accuracy}
nDCGs of TwinBERT$_{\textrm{cos}}$ and C-DSSM at different positions were presented in Figure \ref{fig:retrieval}. The solid lines give the nDCGs at different positions for TwinBERT$_{\textrm{cos}}$ and C-DSSM models. At all positions, TwinBERT$_{\textrm{cos}}$ is consistently better than C-DSSM by at least {5.3\%}. The dashed lines show another group of nDCGs by converting 4-level labels back to binary label. Similarly, TwinBERT$_{\textrm{cos}}$ outperforms C-DSSM by at least {3.6\%}. Both results indicate that TwinBERT$_{\textrm{cos}}$ embeddings capture more information about query and keyword. Table \ref{tab:retrieval} gives the density differences of all 4 labels by comparing top 5 results from both models. The density differences show TwinBERT$_{\textrm{cos}}$ recalls $18.8\%$ more excellent and 7.4\% less bad query-keywords, which proves its superiority in retrieval tasks.

\begin{table}
  \caption{ROC-AUC of TwinBERT w/ actual label fine-tuning (FT) and asymmetric training (ASYM) on two test sets}
  \label{tab:trainingstrategies}
  \begin{tabular}{llll}
    \toprule
    Model & AUC$_{1}$ & AUC$_{2}$\\
    \midrule
    TwinBERT$_{\textrm{res}}$ & 0.9010 & 0.9113 \\
    TwinBERT$_{\textrm{cos}}$ & 0.8883 & 0.8743 \\
    TwinBERT$_{\textrm{res}}$ w/ FT & 0.9030 & 0.9140 \\
    TwinBERT$_{\textrm{cos}}$ w/ FT & 0.8926 & 0.8953 \\
    TwinBERT$_{\textrm{res}}$ w/ ASYM+FT & 0.9033 & 0.9127 \\
    TwinBERT$_{\textrm{cos}}$ w/ ASYM+FT & 0.8982 & 0.9057 \\
  \bottomrule
\end{tabular}
\end{table}
      
\subsection{Effective Training Strategies}
\label{sec:trainingstrategies}
Two training strategies, actual label fine-tuning and asymmetric training, were tested on top of the standard training process and will be discussed in this section independently, as they are orthogonal to the design of TwinBERT model.
\subsubsection{Actual label fine-tuning}
In the standard training process, TwinBERT model learns parameters from soft labels generated by a teacher model. Actual label fine-tuning adds a round of fine-tuning based on the editorial labels post the standard training process. Learning rate was further tuned down to 2e-5. The fine-tuning step took 2 epochs to converge. The first 4 rows of Table \ref{tab:trainingstrategies} give the AUC of TwinBERT models w/ and wo/ actual label fine-tuning on the same test sets used in the relevance experiment. On TwinBERT$_{\textrm{res}}$, the improvements are 0.22\% and 0.3\%, while on TwinBERT$_{\textrm{cos}}$, the improvements are much more significant (0.48\% and 2.4\%). The gains demonstrate positive effect of actual label fine-tuning on both models. Often, a teacher model establishes the upper bound of the performance of student models, and it is worth pointing out that the fine-tuned TwinBERT$_{\textrm{res}}$ has already beat the teacher model on both sets by 0.21\% and 0.03\%, which indicates that by introducing actual labels, the fine-tuning step can bring in additional information. 

\subsubsection{Asymmetric training}
In the current architecture of TwinBERT model, to keep the structure simple, parameters are shared between encoders, while asymmetric parameter training could potentially bring higher performance. To further explore the effect, TwinBERT models were retrained with independent parameters between the encoders. All other training parameters for both standard training and label fine-tuning stayed the same. The last 4 rows of Table \ref{tab:trainingstrategies} give the AUC of models w/ and wo/ asymmetric training. On TwinBERT$_{\textrm{cos}}$, asymmetric training brings 0.63\% and 1.2\% AUC gains on two test sets, which means TwinBERT$_{\textrm{cos}}$ does benefit from the more complex configuration. However, on TwinBERT$_\textrm{res}$, even though training loss has slightly drop, AUC shows +0.03\% and -0.14\% differences on two test sets suggesting asymmetric is barely effective when crossing layer is more complex. 

\subsection{Overall Results}
\label{sec:overallresults}
In summary, the best TwinBERT model (TwinBERT$_\textrm{res}$) achieves 3.4\% and 0.17\% AUC improvements over C-DSSM and BERT$_{3}$ student models following the teacher-student framework, which demonstrates its effectiveness in distilling knowledge from a BERT-like teacher model. On top of that, actual label fine-tuning and asymmetric training help boosting the performance with another 0.26\% incremental gain. \textbf{Overall, the best TwinBERT model outperforms C-DSSM and BERT$_{3}$ models by 3.7\% and 0.42\% while also beats the teacher model, BERT$_{12}$, by 0.24\%}.

\subsection{Inference Time}
\label{sec:inferencetime}
To test the inference time, we implemented TwinBERT and the baseline models using PyTorch based on \cite{Wolf2019HuggingFacesTS} and ran benchmarks on a workstation with the following configuration: Intel\textregistered \space Core\texttrademark\space i7-4790 CPU @ 3.6GHz and 32.0GB memory. To eliminate the impact of noise on queries, we evaluated the average inference time on 1,000 queries and the results are summarized in Table \ref{tab:inf}.

One of the benefits of TwinBERT compared to BERT is that the two inputs are decoupled and if the query stays the same, there is no need to regenerate the query embedding. This could be more clearly explained by the time complexity of TwinBERT and BERT w.r.t number of queries ($N_{q}$) and number of keywords ($N_{k}$). The time complexity of TwinBERT is $O(T_{e}N_{q}(1 + N_{k}) + T_{c}N_{q}N_{k})$, while it is $O(T_{B}N_{q}N_{k})$ for a BERT model. Here, we use $T_{e}$, $T_{c}$, $T_{B}$ to denote the time cost of a single encoder in TwinBERT model, the crossing layer in TwinBERT model and BERT model respectively. Another benefit of TwinBERT is that, in certain scenarios like sponsored search, the keyword embeddings could be pre-computed and loaded in memory so there's no computation for keyword encoding at run-time. Thus, the time complexity of TwinBERT during serving could be even simplified to $O(T_{e}N_{q} + T_{c}N_{q}N_{k})$. In Table \ref{tab:inf}, QEL refers to the number of query encoding loops and the boolean factor (Memory) indicates if the keyword embeddings are in memory. 

The number of keywords is another important factor to consider when talking about efficiency for the initial retrieval phase and refinement/ranking phase after. More specifically, it impacts the time complexity of the evaluation of crossing layers in TwinBERT model $O(T_{c}N_{q}N_{k})$ and the evaluation of BERT model $O(T_{B}N_{q}N_{k})$. In the test, the average number of keywords per query is designed to be 100.

The first two rows in Table \ref{tab:inf} correspond to the inference time for TwinBERT$_{\textrm{cos}}$ and TwinBERT$_{\textrm{res}}$ to score 100 keywords per query assuming only the query-side encoder and crossing layer are performed. The computation cost of cosine similarity is much lower than residual network, which leads to the 8ms difference. Compared to the last two rows of BERT$_3$ and BERT$_{12}$, where the query and keyword are concatenated and encoded as a whole, the efficiency of TwinBERT$_{\textrm{res}}$ is 77 and 422 times faster. With TwinBERT$_{\textrm{cos}}$, the efficiency is even 121 and 663 times faster. Moreover, if the keyword embeddings are generated at run-time, the overall inference time is still better than BERT$_3$ according to Row 3 and Row 5. However, if the query embedding is also repeatedly generated, the inference time of TwinBERT becomes higher than BERT$_3$ as listed in Row 4.

\begin{table}
  \caption{Average inference time for TwinBERT, BERT$_3$ and BERT$_{12}$ over 1,000 queries. QEL refers to the number of query encoding loops. }
  \label{tab:inf}
  \begin{tabular}{llll}
    \toprule
    Model & QEL & Memory & Inf. time (ms)\\
    \midrule
    TwinBERT$_{\textrm{cos}}$ & 1 & $\surd$ & 14 \\
    TwinBERT$_{\textrm{res}}$ & 1 & $\surd$ & 22 \\
    TwinBERT$_{\textrm{res}}$ & 1 & & 1,077 \\
    TwinBERT$_{\textrm{res}}$ & 100 & & 2,144 \\
    BERT$_{3}$ & 100 &  & 1,699 \\
    BERT$_{12}$ & 100 &  & 9,282 \\
  \bottomrule
\end{tabular}
\end{table}

\section{TwinBERT in Production System}
\label{sec:prod}
TwinBERT models have been successfully deployed in the backend of a major sponsored search system and are proved to be very effective and efficient in both retrieval and relevance tasks with acceptable latency. The models achieved 90+\% of the incremental gains observed from a fine-tuned BERT$_{12}$ model and bad ads in online impressions decreased by 10+\% in production. However, the additional serving cost is minimum even on CPUs, compared to the service of a BERT$_{12}$ model which is not feasible on CPUs but requires hundreds of GPUs. On query side, model is served with onnx runtime online and the latency of TwinBERT inference is less than 10ms on average, which could be shadowed by other serving components if it is not on a critical path. On document side, embeddings are prepared offline and indexed in a distributed database for serving. At run-time, only the latency of crossing layers is introduced to the overall latency, which is subtle in a distributed serving system. In addition, the embeddings could be mapped to a lower dimension to further improve the efficiency of crossing layers and also reduce the cost on storage in practise.

\section{Conclusion and Future Work}
\label{sec:con}
The TwinBERT model presented in this paper successfully adapts the technical advances from the pre-trained language models to the area of information retrieval. Decoupling the two inputs and pre-computing embeddings offline improve efficiency by 77+ and 422+ times on BERT$_3$ and BERT$_{12}$ on a presumable number of 100 queries, which enables real-time online serving on CPUs. The innovations on network layers manage to keep majority of the performance gain from BERT$_{12}$, which makes TwinBERT effective in both retrieval and relevance tasks. 


TwinBERT models have demonstrated to be as effective as BERT-Base model. Going forward, more experiments need to be conducted to evaluate the performance with a teacher model that has larger capacity such as BERT-Large. Furthermore, TwinBERT models are developed based on original Transformer. As the increasing demand of model performance, further improvement could be achieved through innovations on Transformer.

To make the presentation pragmatic and intuitive, TwinBERT is introduced in the context of information retrieval. However, TwinBERT is not constrained by a specific problem domain. Looking forward, efforts will be spent on other domains such as question answering.

\bibliographystyle{ACM-Reference-Format}
\bibliography{twinbert}

\end{document}